\documentstyle[psfig]{aipproc}

\def\etal{{et\,al.}}
\def\dashdot{\cdots\cdots\cdots\cdots\cdots\cdots\cdots\cdots\cdots\cdots\cdots
\cdots\cdots\cdots\cdots\cdots\cdots\cdots\cdots\cdots\cdots\cdots\cdots
\cdots\cdots\cdots\cdots\cdots\cdots\cdots}
\def\amin{\ifmmode ^{\prime}\else$^{\prime}$\fi}
\def\asec{\ifmmode ^{\prime\prime}\else$^{\prime\prime}$\fi}
\def\degs{\ifmmode ^{\circ}\else$^{\circ}$\fi}
\def\grad{$^\circ$}
\def\fss{\hbox{$.\!\!^{\rm s}$}}        % Fractions of seconds
\def\h{$^{\rm h}$}\def\m{$^{\rm m}$}

\begin{document}

\title{Rapid ROSAT Observations of small Gamma-Ray Burst Error Boxes}

\author{Jochen Greiner}

\address{Astrophysical Institute Potsdam, 14482 Potsdam, 
Germany}

\maketitle

\begin{abstract}
I summarize the use of the ROSAT satellite over the last months for the 
investigation of well localized GRB error boxes. In 
particular, I  report about 
%(i) a systematic study of several dozens of IPN 
%locations using the ROSAT All-Sky-Survey data, 
 the attempts for and 
results of quick follow-up observations  (including the GRBs 
localized with BeppoSAX and RXTE) with ROSAT with its capability to achieve
10\asec\ locations of X-ray afterglow sources. 
%(ii) some results of deep ROSAT 
%pointings of selected small GRB error boxes on timescales of weeks to
%months after the GRB and 
%(iii) results of the correlation of GRB locations 
%with serendipitous pointed ROSAT observations over the last 
%6 years with particular emphasis on the search for X-ray afterglows.
\end{abstract}

\section*{Introduction}

The discovery of X-ray afterglow emission with the narrow-field instruments
of the BeppoSAX satellite (Costa \etal\ 1997) has
given a dramatic boost to both, observations and theoretical investigations of 
gamma-ray bursts (GRBs) over the last few months. In particular, this finding
has triggered a wealth of rapid follow-up observations in the X-ray, 
optical, infrared and radio bands. This has even lead to the development
of a strategy of localizing the X-ray afterglow of strong GRBs with 
non-imaging X-ray instruments (scanning with the PCA onboard RXTE). 
Also, a procedure has been set up to recognize when individual dwells 
of the RXTE all-sky monitor cover BATSE burst locations and 
to derive arcmin size locations within a few hours (Smith \etal\ 1998).

X-ray follow up observations are primarily performed with the ASCA and ROSAT
satellites. While ASCA's unique strengths are the very rapid response time
(1--2 days) and the high spectral resolution of its X-ray CCD camera 
(Murakami \etal\ 1998), ROSAT can provide arcsec positioning and, due to its 
superb sensitivity, a rather long lever arm for studying the X-ray intensity 
decay up to 14 days after a GRB.
Here I sumarize some results which have been obtained over the last months
with rapid ROSAT follow-up observations of well localized GRB error boxes.
Other areas of GRB research with ROSAT can be found elsewhere, such as
the investigation of small IPN error boxes using the ROSAT all-sky survey data
in Bo\"er \etal\ (1994), a summary of all ROSAT pointings on GRB error boxes 
and a correlation of GRB locations with serendipituous ROSAT pointings in 
Greiner (1997).

\section*{Rapid ROSAT follow-up observations}

Motivated by the occurrence of a few very long lasting GRBs and by the
detection of distinct spectral softening over the burst duration several 
attempts have been made in the past to observe well-localized GRBs 
with ROSAT as quick as possible after the GRB event in the hope to find the 
``smoking gun''. To this end, the GRB had to be localized accurately (to 
fit the field of view) and quickly,
and the GRB location had to be within the
ROSAT observing window ($\approx$30\% of the sky at any moment). 
In retrospective, these ROSAT observations were done not quick enough 
and/or with too short exposure times.

Beginning with GRB 970228 and the establishment of the approximate 1/t
X-ray intensity decay law, ROSAT observations were sensitive enough to detect 
the X-ray afterglow of (some) GRBs. Since then, 
every GRB with a location inside the ROSAT 
viewing zone has been observed with ROSAT as quick as possible. 
We note, that GRBs 970402 and 970508
were not observable for ROSAT due to solar angle constraints. In several
cases, even a second follow-up observation has been performed to allow
a variability check of the detected X-ray sources. Tab. \ref{toos} lists
the GRBs which have been observed as TOO together with the time delay
of the ROSAT observation.

%The fastest response with ROSAT so far is 5 days which is near the minimum
%possible time achievable due to the various scheduling constraints (a curious
%exception is the TOO towards SGR 1806--20 which was triggered by one of the 
%repeating bursts, and turned out to happen just one hour after another
%repeating burst from this source).

   \begin{table}
     \caption{ROSAT target-of-opportunity observations (TOOs) towards 
        GRB locations}
%     \smallskip
     \begin{tabular}{crccccl}
  GRB & Exposure & Detector$^{a)}\!\!$ & Date & Delay & N$_{\rm X}^{b)}$ 
                                                               & Ref.$^{c)}$\\
       &  (sec)~~    &            &  &    &             &             \\
       \noalign{\smallskip}
      \hline
      \noalign{\smallskip}
   920501         &  2784~~~ & P&May 19--20, 1992 & 18 days   & 1  & ~1 \\
   920711         &  2432~~~ & H&Feb. 20--21, 1993 & 28 weeks  & 0  & ~2, 7 \\
 930704/940301   & 3150/1385 & P&Apr. 1, 1994 & ~4 weeks  & 25 & ~3 \\
   960720         &  6960~~~ & H&Aug. 30--31, 1996 & ~6 weeks  & 1  & ~4, 5 \\
   960720         &  2791~~~ & H&Jan. 14--17, 1997 & 24 weeks  & 1  & ~7 \\
   961027-29     & 2065/2499 & P&Feb. 24--28, 1997 & 13 weeks &  54  & ~7  \\
   970111         &  1198~~~ & H&Jan. 16, 1997     & ~5 days   & 0  & ~6 \\
   970111         &   777~~~ & P&Feb. 18, 1997     & ~5 weeks   &  0  & ~7  \\
   970228         & 34280~~~ & H&Mar. 10--13, 1997 & 11--14 days & 1/A & ~8 \\
   970616         & 21950~~~ & H&Jun. 23--25, 1997 & 7--9 days   & 2   & ~9 \\
   970815         & 17115~~~ & H&Aug. 20--22, 1997 & 5--7 days & 1/A & 10  \\
   970828         & 61270~~~ & H& Sep. 3--5, 1997 & 6--8 days   & 1/A   & 11 \\
   971024 & $\approx$15000~~~ & H &  Oct. 30, 1997   & 6 days  &    &  12 \\
   \noalign{\smallskip}
   \noalign{\hspace*{0.15cm}$\dashdot$}
%   \noalign{$\dashdot$}
   \noalign{\smallskip}
   SGR 1806--20 & 1416~~~ & P &Oct. 9--10, 1993 &  12 days (1hr) & 0 & ~7 \\
   SGR 1814--14 & $\approx$10000~~~ & H & Sep. 25--27, 1997 & 13--15 days & 
                                                                        & 12\\
   \noalign{\smallskip}
   \end{tabular}
   \label{toos}

   \noindent\small $^{a)}$ P and H denote PSPC and HRI, respectively. \\
                   $^{b)}$ Numbers of X-ray sources within the GRB error box;
                           A marks afterglow detections. \\
                   $^{c)}$ (1) Hurley \etal (1996), 
                     (2) PI: Hurley,          (3) Greiner \etal\ (1997a),
                     (4) Greiner \etal\ (1996b), (5) Greiner \& Heise (1997)
                     (6) Frontera \etal\ (1997a), (7) Greiner (unpubl.), 
                     (8) Frontera \etal\ (1997b), (9) Greiner \etal\ (1997b),
                     (10) Greiner (1997b), (11) Greiner \etal\ (1997c), 
                     (12) data not yet available during writing.
   \end{table}

\noindent{\bf GRBs 920501 and 960720:}
One X-ray source was found in each of the two GRBs 920501 and 960720.
In both cases there is no strong evidence that these X-ray sources are
related to the GRB. For GRB 960720 the X-ray intensity during a second
observation in Jan.\,1997 is about
a factor 2 higher than 6 weeks after the GRB. This supports
the identification with the (presumably unrelated) variable radio source 
QSO 1729+491 (Greiner \& Heise 1997).

\begin{figure}[t]
  \vbox{\psfig{figure=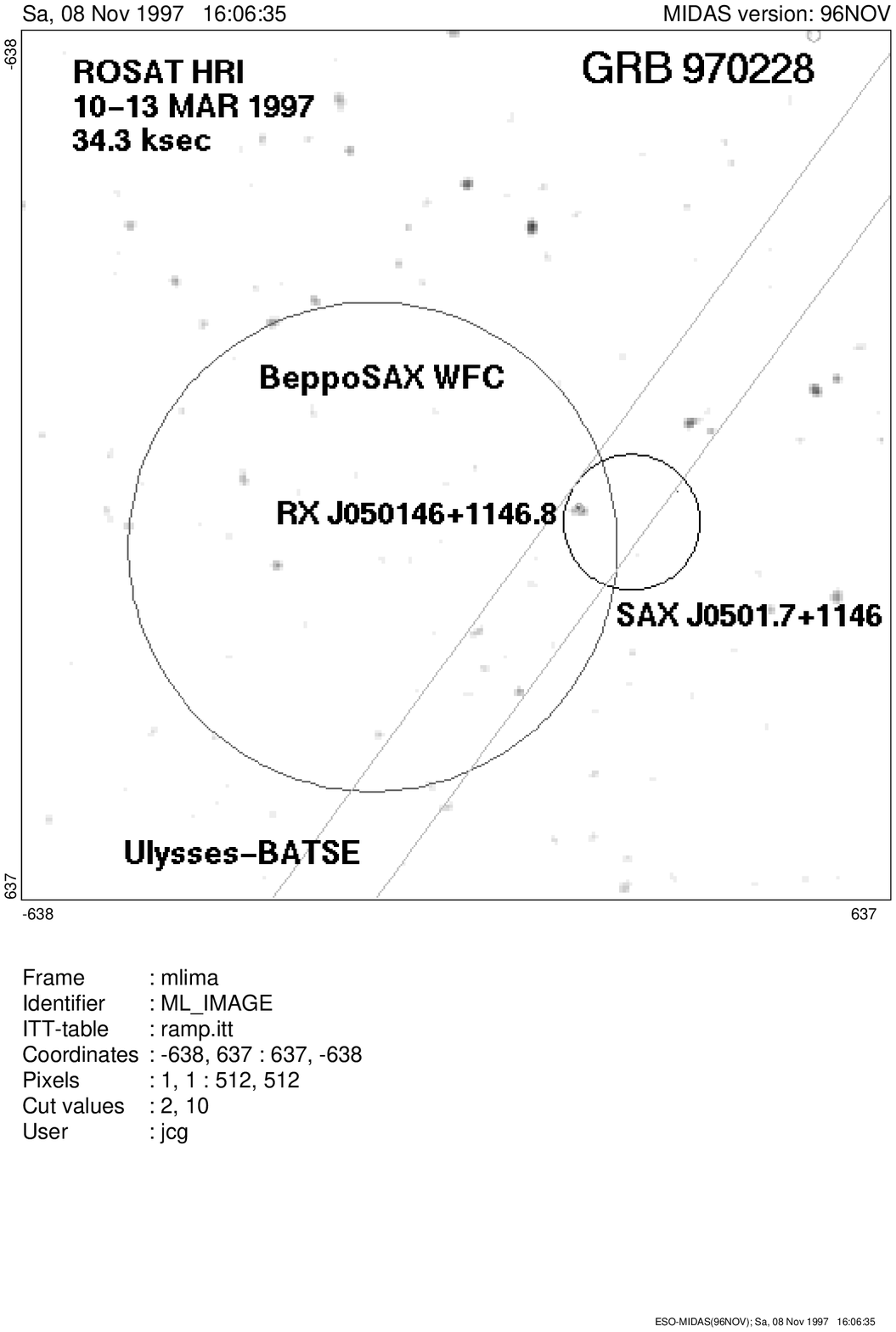,width=0.48\textwidth,%
          bbllx=3.1cm,bblly=10.5cm,bburx=18.7cm,bbury=26.1cm,clip=}}\par
\vspace*{-6.78cm}\hspace*{7.cm}
    \vbox{\psfig{figure=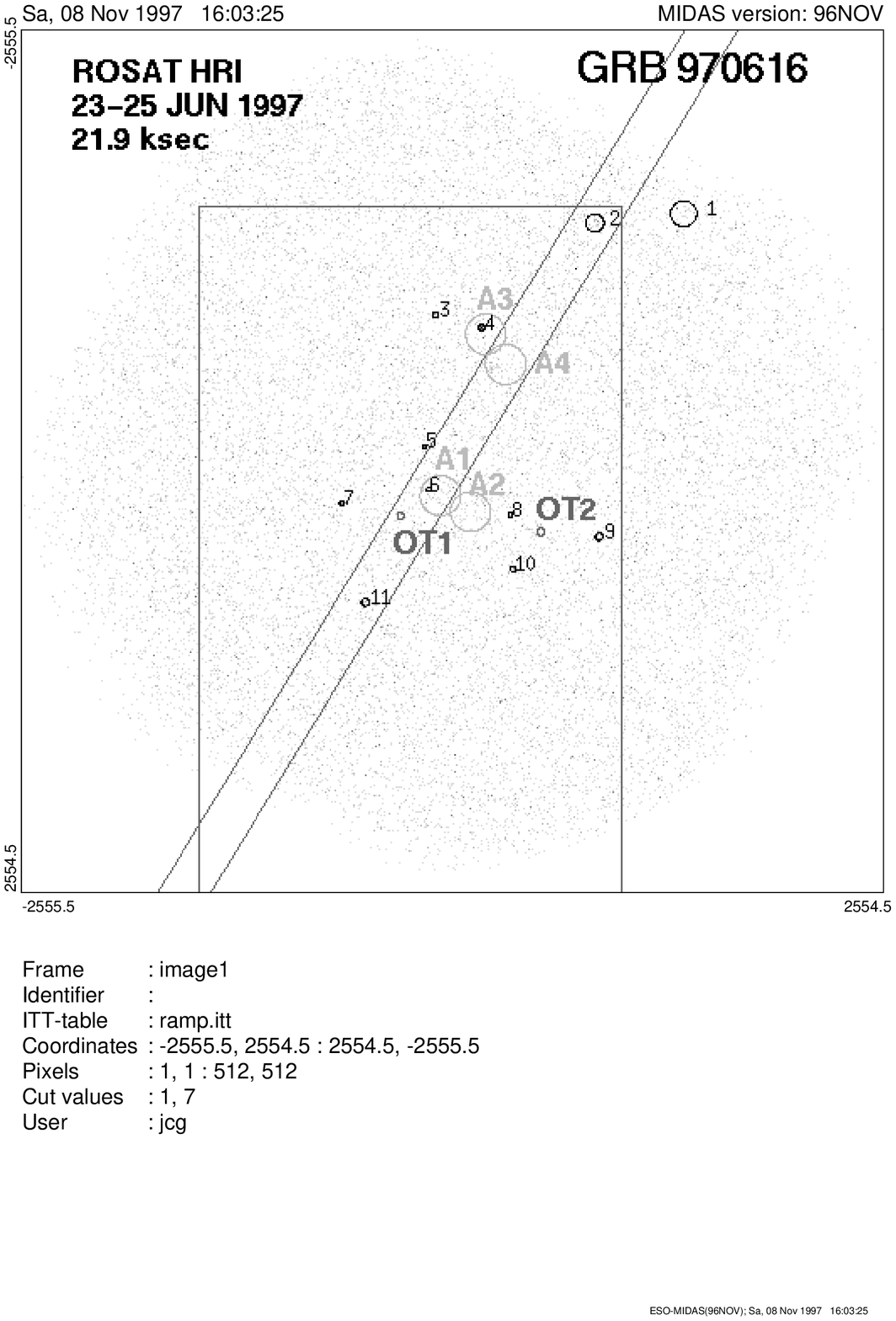,width=0.48\textwidth,%
          bbllx=3.1cm,bblly=10.5cm,bburx=18.7cm,bbury=26.1cm,clip=}}\par
  \caption[ros0228]{{\bf Left:} 
  Central 8\amin\ part of the HRI FOV of the ROSAT observation of GRB 970228.
  The only source above 3$\sigma$ is RX J050146+1146.8.
  The large circle shows the 3$\sigma$ WFC error circle of GRB 970228, 
  the small circle is the $\approx$1\amin\ position 
  of the fading source SAX J0501.7+1146 (BSAX NFI 
  pointings) and the two straight lines mark the triangulation arc derived
  from the BATSE/Ulysses timings  (Hurley \etal\ 1997a, Cline 
  et al 1997). (Adopted from Frontera \etal\ 1997c) 
  {\bf Right:}     Image of the ROSAT HRI observation of GRB 970616: The
  large square denotes the  RXTE error box, the straight lines are the 
  Ulysses/BATSE triangulation arc (Hurley \etal\ 1997b), the enumerated
  circles denote the ROSAT X-ray sources (Greiner \etal\ 1997b), the
  circles marked with A1--A4 are the 4 ASCA X-ray sources detected 4 days
  after the GRB (Murakami \etal 1997a) and
  OT1/OT2 denote the two variable optical objects reported 
  (Galama \etal\ 1997, Udalski 1997).  
  Colour versions of these images are available at
   http:/$\!$/www.aip.de:8080/\~\,jcg/grbgen.html.
\label{rosfig}
}
\end{figure}

\noindent{\bf GRB 970228:} 
The last glimpse of the X-ray afterglow of GRB 970228 was captured with ROSAT
(Fig. \ref{rosfig}, left) at 11--14 days after the burst. Its 
 position with an error radius of 10\asec\ is well within the combined 
WFC/IPN error box and  coincides within 
2\asec\  with the optical transient (Paradijs \etal\ 1997). 
The measured X-ray count rate during the ROSAT observation is consistent with 
an X-ray intensity decay law of t$^{-1.25\pm0.15}$ (Frontera \etal\ 1997c).

\noindent{\bf GRB 970616:}
For  GRB 970616 (Fig. \ref{rosfig}, right) three X-ray sources 
have been found with ROSAT in the RXTE/IPN box,
one of which (R.A. = 01\h18\m50\fss3, Decl. = --05\degs25\amin58\asec)
coincides with one of the 4 X-ray sources found with ASCA (Murakami \etal\ 
1997a). The unabsorbed ROSAT flux (using a power law model with photon index 
2 and the galactic absorbing column N$_{\rm H} = 4\times 10^{20}$ cm$^{-2}$) 
is 1.4$\times$10$^{-14}$
erg/cm$^2$/s in the 0.5--2 keV range, i.e. a factor 5 lower than the flux
measured by ASCA on June 20/21 (using the same model assumptions and energy 
range). The 3$\sigma$ upper limit of the flux at the position of the ASCA 
source A4 is a factor 7 lower 
than that of the ASCA detection. With  two fading X-ray
sources within the RXTE/IPN error box an 
 association to the GRB remains open.

\noindent{\bf GRB 970815:}
One faint X-ray source is detected with ROSAT at R.A. = 16\h08\m47\fss7, 
Decl. = +81\degs31\amin53\asec\ (equinox 2000.0), which just falls
on the border of the RXTE error box (Greiner 1997b).
Using a power-law model with photon index 2 and the galactic absorbing 
column N$_{\rm H}$=4.7$\times$10$^{20}$ cm$^{-2}$, the observed count 
rate of 0.00073$\pm$0.00025 cts/s
corresponds to an unabsorbed ROSAT flux (0.1--2.4 keV) of 
5$\times$10$^{-14}$ erg/cm$^2$/s. This flux level is consistent with a
t$^{-1}$ flux decay, suggesting that it may be the afterglow of GRB 970815. 

\noindent{\bf GRB 970828:}
One X-ray source, located at R.A. = 18\h08\m31\fss7, 
Decl. = +59\degs18\amin50\asec\ (equinox
2000.0; uncertainty $\pm$10\asec; RX J1808.5+5918), is detected in the 
longest ever ROSAT TOO pointing within the RXTE/IPN error box and 
coincides with the X-ray source detected with ASCA 5 days earlier 
(Murakami \etal\ 1997b).  Using a power-law model with photon index 2
and the galactic column N$_{\rm H}=3.7\times$10$^{20}$ cm$^{-2}$, the observed
countrate of (4$\pm$1)$\times$10$^{-4}$ cts/s corresponds to an unabsorbed 
ROSAT flux (0.1--2.4 keV) of 2.5$\times$10$^{-14}$ erg/cm$^2$/s, 
consistent with a power-law decay.

%While originally the discovery of
%a quiescent X-ray source inside a small GRB error box has been considered as
%probable evidence for an association of a GRB with a quiescent counterpart,
%the continuing discovery of further X-ray sources and in particular the
%detection of more than one X-ray source even in small GRB error boxes
%as well as the fading afterglow emission for the recent GRBs 
% makes this association doubtful. Also, the optical
%identification of these X-ray sources, though not yet completely established 
%in all cases, does not find evidence for unusual objects.

\noindent{\bf Density of background (foreground?) X-ray sources:}
%An estimate of the chance probability for the occurrence of a quiescent
%X-ray source inside a well-localized GRB error box depends on how the 
%question is asked in detail (see Hurley \etal\ 1996 for various 
%possibilities). However, one has to bear in mind that 
At the low sensitivity limits reached during the rather long pointings, 
the number density of X-ray sources is already remarkably high. From 
the results of many deep pointed observations and combined with the
particularly deep Lockman hole observations of ROSAT an improved log\,N--log\,S
distribution of X-ray sources has been derived (Hasinger 1997) which
gives 100--700 X-ray sources per 1\,$\Box$\grad\, 
 at the level of 10$^{-14}$...10$^{-15}$ erg/cm$^2$/s.
Thus, the probability for a chance coincidence
of a quiescent, soft X-ray source with a GRB location 
is 25\%--100\% for a 10 arcmin$^2$ size error box.

\section*{Conclusion}

The search for X-ray counterparts (afterglows) in small  error boxes 
re\-mained inconclusive in the past as long as the observation was performed
more than 2 weeks after the GRB. Only the 10\asec\ localization with 
ROSAT (Frontera \etal\ 1997c) of the fading X-ray source within the 
GRB 970228 error box (Costa \etal 1997) has allowed to unequivocally associate
it with the optical transient (Paradijs \etal\ 1997), the properties
of which (location within extended fuzz) have been used to conclude the 
cosmological distance scale of GRBs.

The X-ray afterglows of two further GRBs have been localized with ROSAT, 
but despite the 10\asec\ accuracy no optical transients could be identified
even several magnitudes fainter than the
GRB 970228 OT brightness. While this has been interpreted as
a consequence of absorption inside the GRB host site (discussion between 
B. Paczynski and J. van Paradijs), it emphasizes the need for further 
GRB/afterglow positions at the arcsec level in order to avoid ambiguities
due to background sources (in the X-ray as well as optical bands).

%\vspace{0.1cm} 
\noindent{\small\it Acknowledgements: 
I'm highly indebted to the SAX team, particularly L.\,Piro, J.\,Heise, 
E.\,Costa,
F.\,Frontera), the BATSE PCA team and the RXTE/ASM team, most notably
D.A.\,Smith and R.A.\,Remillard for the rapid notification of GRB positions.
It is a great pleasure to thank J. Tr\"umper for granting substantial ROSAT 
time for target-of-opportunity observations of GRB error boxes.
I also greatly acknowledge the substantial contributions of  MPE members
(among them W.\,Voges, Th.\,Boller, J.\,Englhauser and J.\,Siebert) at the
various stages of the data acquisition and processing.
%I'm indebted to J. Englhauser (MPE) for the help in handling the OBI database.
I appreciate substantial travel support from the conference organizers. 
The author is supported by the German Bundesministerium f\"ur Bildung,
Wissenschaft und Forschung (BMBW/DARA) under contract No. 50 QQ 9602 3. 
The ROSAT project is supported by the BMBW/DARA and the Max-Planck-Society.}

\end{document}